\documentstyle[12pt]{article}
\begin{document}
\newcommand{\nn}{\noindent}
\newcommand{\nll}{\nonumber \\}
\newcommand{\hf}{\hfill}
\newcommand{\bq}{\begin{equation}}
\newcommand{\eq}{\end{equation}}
\newcommand{\bea}{\begin{eqnarray}}
\newcommand{\eea}{\end{eqnarray}}
\newcommand{\nobody}{\rule{0ex}{1ex}}
\newcommand{\nobodyfrac}{\frac{\nobody}{\nobody}}
\newcommand{\hnobody}{\rule{1ex}{0ex}}
\newcommand{\ra}{\rightarrow}
\begin{titlepage}
\vskip 35pt
\large
\centerline {\bf On the energy dependence of hadronic $B_c$ production
  \footnote{Contribution to the 5th International Workshop on $B$-Physics
            at Hadron Machines, 1997, Los Angeles, California, USA}$^,$
  \footnote{Work supported by the Bundesministerium f\"ur Bildung, 
            Wissenschaft, Forschung und Technologie, Bonn, Germany,
            Contract 05 7WZ91P(0).}} 
\normalsize
 
\vskip 2.0cm
\centerline {K.\ Ko\l odziej\footnote{On leave of 
            absence from the Institute of Physics, University of Silesia, 
            PL-40007 Katowice, Poland}
            and R.\ R\"uckl}
\vspace{3.0mm}
\centerline {\it Institut f\"ur Theoretische Physik, Universit\"at 
W\"urzburg}
\centerline{\it D-97074 W\"urzburg, Germany}
\vskip 3.5cm
\centerline {\bf Abstract}
\vskip 1.0cm 
An estimate is presented of the 
production cross section of $B_c$ mesons between threshold and
LHC energies using lowest-order perturbation theory and 
non-relativistic bound state approximation. It is shown that the ratio
of the production cross sections for $B_c$ mesons and for $b$ quarks varies
strongly with energy.
\vfill
\end{titlepage}

The results of References~[1-3] are sometimes used to
estimate the production rate of B$_c$ mesons at lower energies 
than they were intended for.
In Ref. \cite{lusignoli}
the HERWIG parton shower Monte Carlo code is used to determine 
the ratio of the $B_c$ and $b\bar{b}$ cross sections, 
while in Refs. \cite{cheung,amiri}
the probability for a $\bar{b}$
quark to fragment into a $B_c$ meson is calculated perturbatively 
in the nonrelativistic bound state approximation.
Both estimates suggest a probability
of order $10^{-3}$ for $b$-quark 
fragmentation into $B_c$ mesons relative to fragmentation into $B$ mesons.

By direct perturbative calculation \cite{klr2,CC,BLS}
of the complete reaction, 
$pp$ (and $p\bar{p}$) $\rightarrow B_c b \bar{c} X$, 
it has been shown that at high energies and large transverse momenta,
hadronic $B_c$ production   
is indeed well described by $b$-quark fragmentation.
However, at low energies near threshold and at small $B_c$ transverse momenta,
both $\bar{b}$-quark fragmentation and $\bar{b}c$ recombination
processes are expected to contribute. 
This puts some doubt on the validity of the fragmentation description
\begin{equation}
\label{frag}
d \hat{\sigma}(B_c) =
d \hat{\sigma}(b\bar b) \otimes D_{\bar b \rightarrow B_c\bar c} (z)
\end{equation}
for estimates of total production rates.
Here, the familiar short-hand notation for the convolution 
integral over the fragmentation function
$D_{\bar b \rightarrow B_c\bar c}(z)$ is used.
To clarify the issue, we have calculated the production rates 
of $B_c$ mesons focussing on the threshold region. 

At low energy, two subprocesses contribute to the production of $B_c$ mesons:
\bea
\label{ggbc}
g g & \rightarrow & B_c b \bar c, \\
\label{qqbc}
q \bar q & \rightarrow & B_c b \bar c, \qquad q = u, d, s.
\eea
At larger energies gluon-gluon fusion dominates so that the $pp$ and 
$p \bar{p}$ cross sections become equal.
In leading order, one has to calculate the $O(\alpha_s^4)$ hard scattering
amplitudes for the $2 \rightarrow 4$ processes  
$gg$ (and $q \bar{q}$) $\rightarrow b \bar{b} c \bar{c}$, and convolute them
with the wave function describing the bound state formation
$\bar{b} c \rightarrow B_c$.
In the nonrelativistic approximation, the binding energy 
and relative momentum of the $\bar b$- and $c$-quarks are neglected.
Moreover, the product of the $\bar b$- and $c$-quark spinors appearing
in the amplitudes of $b \bar b c \bar c$ production are substituted
by spin and energy projectors \cite{guberina}. 
In particular, for the pseudoscalar and vector ground states one has
\begin{equation}
\label{subst}
v(p_{\bar b})\bar u(p_c) = \frac{f_{B_c^{(*)}}}{\sqrt{48}}(\not{p} - m_{B_c})
                           \Pi_{SS_Z} ,
\end{equation}
where the spin projector $\Pi_{SS_Z}$ is equal to $\gamma_5$ for the 
$B_c$ and to $/\!\!\!{\varepsilon}$ for the $B_c^*$.
The decay constants $f_{B_c^{(*)}}$ being related to the S-wave function
at the origin, can be estimated in potential models or by QCD sum rules.
Finally, the cross sections of the subprocesses 
(\ref{ggbc}) and (\ref{qqbc}) are to be folded with the parton densities 
of the proton.
A detailed description of the calculation for the gluon-fusion process
(\ref{ggbc}) can be found in \cite{klr2}. 

Taking $m_b=4.8$ GeV, $m_c=1.5$ GeV, $m_{B_c^{(*)}}=6.3$ GeV and 
$f_{B_c^{(*)}}=0.4$ GeV \cite{rr}, and using the parametrization 
of structure functions from Ref. \cite{mrs},
we get, at the HERA--B energy of $\sqrt{s} = 40$~GeV:
\begin{equation}
\label{cross}
\sigma (B_c) = 2.5 \, {\rm fb}, 
\end{equation}
\begin{equation}
\label{cross2}
\sigma (B_c^*) = 8.3 \, {\rm fb}.
\end{equation}
Here we have used the running coupling constant 
$\alpha_s(\mu^2)$ in leading logarithmic 
approximation for five flavours and with the fixed scale  
$\mu^2 = 4m_{B_c}^2$. As expected, the scale-dependence of the prediction
is very strong. For illustration, the above cross sections
shrink by a factor 3
when choosing $\mu^2 = 4x_1x_2s$ and increase by a factor 10 
when taking $\mu^2 = {1\over 4}(p_T^2+m_{B_c}^2)$.
Because of this huge uncertainty which can only be reduced by 
including higher order corrections, it is more sensible to
consider the cross section ratio $\sigma (B_c + B_c^*) / \sigma (b \bar{b})$,
where some of the scale dependence can be expected to cancel.
The ratio is shown in Fig.~1 as a function of the c.m. energy.
We see that the theoretical uncertainty due to the scale ambiguity is
indeed reduced to a factor 3 to 4.
In particular, at $\sqrt{s} = 40$ GeV, we predict:
\begin{equation}
\label{ratio}
{{\sigma (B_c+B_c^*)}\over {\sigma (b\bar{b})}} = 0.35 \, (1.2) 
            \times 10^{-5},
\end{equation}
for $\mu^2 = 4x_1x_2s$ ($\mu^2 = {1\over 4}(p_T^2+m_{B_c}^2)$).
The ratio rises to about $10^{-4}$ at Tevatron energies and 
continues to increase slowly as one approaches the LHC energy.  
It should be noted that uncertainties due to the quark masses,
decay constants, etc. are not included in the above estimates. 

Taking into account that not all $b$-quarks lead to $B$-mesons and
that there are heavier $\bar{b} c$ bound states with masses below the 
$BD$ production threshold which decay into the  
$B_c$ ground state, the ratio $\sigma (B_c) / \sigma (B)$ 
of inclusive cross sections
may actually reach the value $10^{-4}$ at HERA--B and $10^{-3}$
at Tevatron energies.
A more detailed investigation
shows that the fragmentation description fails 
at $p_T \le 5$ GeV, leading to an overestimate of the integrated 
$B_c$ cross section, in particular in the threshold region.

\pagebreak

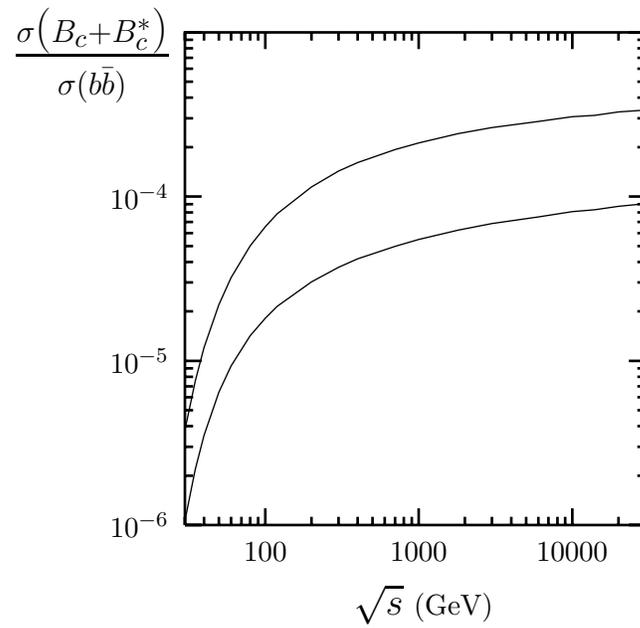
\begin{figure}
\centering
%
\setlength{\unitlength}{0.1bp}
\begin{picture}(2519,2160)(0,0)
\put(1500,-49){\makebox(0,0){{\large $\sqrt{s}$} {\small (GeV)}}}
\put(250,1980){\makebox(0,0)[b]{\shortstack{{\Large ${{\sigma \big( B_c+B_c^*\big)}                  \over{{\rule[-1mm]{0mm}{6mm}\sigma ( b\bar b)}}}$}}}}
\put(2060,151){\makebox(0,0){{\small $10000$}}}
\put(1481,151){\makebox(0,0){{\small $1000$}}}
\put(903,151){\makebox(0,0){{\small $100$}}}
\put(540,1490){\makebox(0,0)[r]{{\small $10^{-4}$}}}
\put(540,870){\makebox(0,0)[r]{{\small $10^{-5}$}}}
\put(540,251){\makebox(0,0)[r]{{\small $10^{-6}$}}}
\end{picture}
\caption{The ratio of the production cross sections 
for $B_c$ mesons and $b$ quarks as a function of the proton--proton 
c.m. energy for $\mu^2 = 4x_1x_2s$ (lower curve) and 
$\mu^2 = {1\over 4}(p_T^2+m_{B_c}^2)$ (upper curve).}
\label{fig:one}
\end{figure}

\end{document}